\documentclass[runningheads]{llncs}
\newif{\ifanonymized}
\anonymizedfalse
\usepackage[T1]{fontenc}
\usepackage{stmaryrd}
\usepackage{graphicx}
\usepackage{wrapfig2}
\usepackage{amsmath,amsfonts,amssymb}
\usepackage{mathpartir}
\usepackage{tikz-cd}
\usetikzlibrary{fit}
\usepackage{listings}%
\newcommand{\lst}[1]{\text{\lstinline{#1}}}
\usepackage{subcaption}

\ifanonymized
  \def\hyperrefparams{colorlinks,pdftex}
\else
  \def\hyperrefparams{colorlinks,pdftex,pdfauthor={Jun Inoue}}
\fi

\usepackage[\hyperrefparams]{hyperref}
\usepackage{cleveref}
\crefname{appendix}{Appendix}{Appendices}
\crefname{section}{Section}{Sections}
\crefname{figure}{Figure}{Figures}
\crefname{table}{Table}{Tables}

\crefname{axiom}{Axiom}{Axioms}
\crefname{theorem}{Theorem}{Theorems}
\crefname{lemma}{Lemma}{Lemmas}
\crefname{proposition}{Proposition}{Propositions}
\crefname{definition}{Definition}{Definitions}
\crefname{corollary}{Corollary}{Corollaries}
\crefname{example}{Example}{Examples}
\crefname{algorithm}{Algorithm}{Algorithms}
\crefname{assumption}{Assumption}{Assumptions}
\crefname{remark}{Remark}{Remarks}
\crefname{question}{Question}{Questions}
\crefname{idea}{Idea}{Ideas}
\crefname{summary}{Summary}{Summaries}
\crefname{enumi}{}{}
\crefname{enumii}{}{}
\crefname{equation}{}{}

\newcommand{\ketbra}[2]{\left|{#1}\middle\rangle\middle\langle{#2}\right|}
\newcommand{\bracket}[3]{\left\langle{#1}\middle|{#2}\middle|{#3}\right\rangle}

\newcommand{\ket}[1]{\left|#1\right\rangle}
\usepackage{multicol}

\newcommand{\meas}[2]{\operatorname{measure}^{#1}_{#2}}

\newcommand{\sem}[1]{\left\llbracket #1 \right\rrbracket}
\newcommand{\veclen}[1]{\lvert #1 \rvert}

\newcommand{\sqrthalf}{\frac{1}{\sqrt{2}}}

\begin{document}
\title{Quantum Programming Without\\ the Quantum Physics}
%
%
\author{\ifanonymized (anonymized for review) \else Jun Inoue\orcidID{0000-0002-2939-8337}\fi}
\authorrunning{\ifanonymized (anonymized for review)\else J. Inoue\fi}
%
\institute{\ifanonymized \else National Institute of Advanced Industrial Science and Technology \\
  1-8-31 Midorigaoka Ikeda, Osaka, Japan
  \email{jun.inoue@aist.go.jp}\fi}
\maketitle              

%
\begin{abstract}
  We propose a quantum programming paradigm where all data are 
  familiar classical data, and the only non-classical element is a 
  random number generator that can return results with negative 
  probability.
  Currently, the vast majority of quantum programming languages 
  instead work with quantum data types made up of qubits.
  The description of their behavior relies on heavy linear algebra 
  and many interdependent concepts and intuitions from quantum 
  physics, which takes dedicated study to understand.
  We demonstrate that the proposed view of quantum programming 
  explains its central concepts and constraints in more accessible, 
  computationally relevant terms.
  This is achieved by systematically reducing everything to the 
  existence of that negative-probability random generator, avoiding 
  mention of advanced physics.
  This makes quantum programming more accessible to programmers 
  without a deep background in physics or linear algebra.
  The bulk of this paper is written with such an audience in mind.
  As a working vehicle, we lay out a simple quantum programming 
  language under this paradigm, showing that not only can it express 
  all quantum algorithms, it also naturally captures the semantics of 
  measurement without ever mentioning qubits or collapse.


\end{abstract}
\section{Introduction}

In this article, we reconstruct quantum programming in terms that are 
readily intelligible to programmers with only a passing knowledge of 
linear algebra or quantum theory.
The key idea, inspired by Aaronson 
\cite{Aaronson2013QuantumComputingDemocritus} and Mu and Bird 
\cite{MuBird2001FunctionalQuantumProgramming}, is that \emph{quantum 
  programming is probabilistic programming with negative 
  probabilities}, where ``probabilistic programming'' just means 
``the writing of programs that call random number 
generators''.\footnote{In the literature, ``probabilistic 
  programming'' has come to mean a related but broader set of 
  activities, which is discussed in the related works section.}

We show that universal quantum computation (in the sense of 
computational universality \cite{Aharonov2003SimpleProofToffoli}) can 
be achieved by leveraging nothing but a random number generator that 
returns certain results with negative probability.
A negative-probability outcome is like a normal outcome, except that 
it can cancel with a positive-probability outcome that leads to the 
same machine state; canceled outcomes are then never observed.
What emerges is a new programming model, quantum probabilistic 
programming (QPP), which dispenses entirely with the notion of qubit, 
working with familiar classical data.
Quantum weirdness is instead pushed into the computational effect of 
quantum non-determinism.

Of course, probability in the usual sense cannot be negative.
Negative probabilities are probabilities in only a generalized sense, 
but by drawing parallels with ordinary probabilities, this concept 
helps to foster intuitions about the behavior of quantum programs.
Constraints on quantum computation like reversibility are derived 
from commonsense assumptions as to what characterizes probabilistic 
programming, rather than imposed on an ad-hoc basis.
Even measurements can be understood without qubits as effective loss 
of signs on probabilities, again allowing constraints to be explained 
rather than imposed.

Why do this?
The point of the exercise is to rectify the current situation in 
which quantum programming is very difficult to get into.
A dizzying array of quantum programming languages exist 
\cite{SandersZuliani2000QuantumProgramming,%
  vanTonder2003LambdaCalculusQuantum,%
  JorrandLalire2005QuantumPhysicsProgramming,%
  AltenkirchGrattage2005FunctionalQuantumProgramming,%
  LampisEtAl2008QuantumDataControl,%
  SelingerValiron2009QuantumLambdaCalculus,%
  YingEtAl2009AlgebraQuantumProcesses,%
  GreenEtAl2013QuipperScalableQuantum,%
  WeckerSvore2014LIQUiSoftwareDesign,%
  SvoreEtAl2018EnablingScalableQuantum,%
  PechouxEtAl2020QuantumProgrammingInductive,%
  BichselEtAl2020SilqHighlevelQuantum}, but they invariably define a 
qubit type, which denotes a 2-dimensional Hilbert space, arrays of 
which denote tensor products, which are operated on by unitaries.
This is already enough non-computer-related jargon to give the novice 
a run for Wikipedia, but it is only the beginning.
There follows a journey of confusion through intuition-free 
abstractions and unexpected constraints, with many of the latter 
coming out of seemingly random physics facts.
To wit:
\begin{itemize}
\item Qubits' distinguishing feature is their ability to be in 
  superposition, but the meaning of this is already murky.
  Many sources say it means being ``0 and 1 simultaneously'', 
  enabling massive parallelization, but prominent experts warn 
  against this intuition \cite{Aaronson2021WhatMakesQuantum}, 
  offering no good alternatives.
\item Qubits can't be copied due to no-cloning, a result which holds 
  because cloning would violate Heisenberg's uncertainty principle.
\item All operations must be reversible because it must be unitary, 
  which in turn is because of the general form of solutions to 
  Schr\"odinger's equation.
\item Qubits can't be discarded willy-nilly because discarding 
  somehow involves measurement.
  Can't we just throw away qubits without looking at them?
\item Measurement of one qubit modifies other entangled qubits, but 
  this can't be used to send information because of special 
  relativity.
\item Entanglement seems to mean correlation between qubits' values, 
  but initializing qubits to all 0's, giving perfect correlation, 
  somehow doesn't count.
\item Entanglement underlies many counter-intuitive behaviors: for 
  instance, merely referencing qubit A to decide whether to flip 
  qubit B can affect the value of qubit A later on (see \cite[Section 
  3]{ParadisEtAl2021UnqompSynthesizingUncomputation} for an example).
\end{itemize}
While trying to internalize all of this, one must wade through heaps 
of complex-number linear algebra, which is alien to most other kinds 
of programming.


The root of the problem is that quantum computing has been framed in 
terms familiar to physicists and not to computer scientists, let 
alone ordinary programmers.
As such, one has to navigate a sea of concepts and facts that offer 
rich physical meanings but little intuition on how they help or 
constrain program construction.
It is quite telling how textbooks in this field generally start with 
expositions of quantum physics rather than programming models or 
primitives \cite{NielsenChuang2011QuantumComputationQuantum,%
  CoeckeKissinger2017PicturingQuantumProcesses,%
  Hidary2021QuantumComputingApplied}.
Existing quantum programming languages have inherited this 
physics-centric, linear algebra-heavy view, starting with the concept 
of qubits.
As such, until programmers grok a nontrivial amount of quantum 
physics, they can barely have a sense that they've understood the 
principles behind a quantum programming language's design, making it 
seem arbitrary and unpredictable.


In this article, we show that negative probabilities provide a 
simpler, more computationally appealing metaphor for explaining 
quantum programming concepts, facts, and constraints than qubits.
For example:
\begin{itemize}
\item Data inside quantum computers are familiar classical data.
  They do not entangle, and they can be copied.
\item Superposition is just a list of possible outcomes of a 
  randomized trial, weighted by probability.
  It's just that, unlike with classical randomness, the outcomes can 
  interact---they can later meet and cancel if they occur with 
  opposite-sign probabilities.
\item Unitarity is just the law that probabilities always add to 1, 
  in the presence of negative probabilities.
\item Measurement is the loss of signs from some of the 
  probabilities.
  Its irreversibility is helpful in only limited contexts because it 
  merely succeeds in being irreversible at throwing away the 
  advantages of quantum computing.
\item Discarding behaves like measurement because both are about 
  information leakage that ruins the programmer's control over 
  cancellation.
\end{itemize}
In this way, negative probabilities tie together all the concepts 
that a programmer needs to digest in order to get a feel for quantum 
programming.

Though some authors have written on the idea of negative 
probabilities before, it was for different purposes.
Aaronson \cite{Aaronson2004QuantumMechanicsIsland,%
  Aaronson2013QuantumComputingDemocritus} quipped that quantum 
\emph{theory} (not programming) is what one gets by incorporating 
negative numbers to probability theory.
Mu and Bird, in an apparently unpublished draft 
\cite{MuBird2001FunctionalQuantumProgramming}, explored qubit-less 
programming with negative probabilities, trying to understand quantum 
non-determinism as a monad-like structure.
But neither tried to systematically reduce concepts to negative 
probabilities, putting together a comprehensive narrative of how 
quantum programming works, embodied in a concrete programming 
language.
This usage, to our knowledge, is novel.

Our treatment of measurement is especially noteworthy.
We split uncertainty in a quantum program into two layers, a quantum 
layer with signed probabilities and a classical layer with ordinary, 
unsigned probabilities.
We then explain measurement as a movement of uncertainty from the 
quantum to the classical layer---effectively, a loss of signs.
While mathematically just a rehash of ensembles and density matrices, 
this view again offers great simplifying power.
Not only does it avoid discussions of qubits and their intricate 
behavior under measurement, it also clarifies the relationship 
between measurement, discarding, output, and information leakage, all 
while relying on just one difference between quantum and classical 
randomness---signed probabilities.

The contributions of this paper are as follows.
\begin{itemize}
\item We define what it means for classical probabilistic programming 
  (reviewed in \cref{sec:cppl}) to be fitted with negative 
  probabilities, giving a simple quantum programming language QPPL.
  We show how the other features of quantum programming arise from 
  that one modification (\cref{sec:qppl}).
\item Through a simple programming example, we show how negative 
  probabilities help to understand and motivate quantum algorithms 
  (\cref{sec:examples}).
\item We show how QPP offers an understanding of measurement and 
  discarding that does not depend on the concept of qubits.
  By splitting uncertainty into quantum and classical layers, we can 
  formulate them as losses of control over cancellation due to 
  information leakage (\cref{sec:measurements}).
\item We define a formal semantics for QPPL and show that it is both 
  implementable and computationally universal 
  (\cref{sec:formal-semantics}).
\end{itemize}
\Cref{sec:cppl,sec:qppl,sec:examples,sec:measurements} should be 
accessible to readers without a background in quantum computing or 
programming language theory.
A modicum of basic linear algebra over the real numbers is required: 
the concept of linearity, conditions for injectivity of linear maps, 
dot products, and orthogonality.

\section{Background: Classical Probabilistic Programming}
\label{sec:cppl}

In this section, we review how classical probabilistic programming 
and its semantics can be conceptualized, setting up notation and 
vocabulary.

\newsavebox{\myifbox}
\begin{figure}[t]
  \centering
  \[
    \begin{tikzcd}[row sep=1ex,/tikz/column 1/.append style={anchor=base west}]
      \lst{def main(x,y : bit):} &&
      & \ket{00}\ar[ddl,"\frac{1}{2}"']\ar[ddr,"\frac{1}{2}"] \\
      \hspace{2em}\lst{x := rand_bit()}
      \\
      &&
      \frac{1}{2}\ket{00} \ar[dd,"\frac{1}{2}"'] \ar[ddrr,"\frac{1}{2}" pos=5/6]
      & +
      & \frac{1}{2}\ket{10} \ar[ddll,"\frac{1}{2}"' pos=5/6] \ar[dd,"\frac{1}{2}"] \\
      \hspace{2em}\lst{x := rand_bit()}
      \\
      &&
      \frac{1}{2}\ket{00} \ar[dd,"1"'] & + & \frac{1}{2}\ket{10} \ar[dd,"1"]
      \\
      \hspace{2em}\lst{y := x}
      \\
      \hspace{2em}\lst{return x,y}
      &&
      \frac{1}{2}\ket{00} & + & \frac{1}{2}\ket{11}
    \end{tikzcd}
  \]
  \caption{Example of a classical probabilistic program and its 
    transition diagram.
    Each row of nodes in the diagram shows which worlds (i.e.\ 
    states) the program can be in at the source line aligned with 
    that row, as a sum weighted by probabilities.
    Arrows show which worlds transition to which given the source 
    lines next to them, labeled with transition probabilities.}
  \label{fig:eg:cpp}
\end{figure}

\cref{fig:eg:cpp} shows an example of a (contrived) classical 
probabilistic program in a toy language, CPPL.
On the left is a program that declares two inputs \lstinline{x,y} of 
type \lstinline{bit}, both implicitly initialized to 0.
It generates a random bit twice, storing the result in \lstinline{x} 
each time, overwriting the first result by the second.
The second result is copied to \lstinline{y}, and the two copies are 
returned as the final program result.
For simplicity, CPPL only allows \lstinline{return} at the end of the 
program.

The program's source lines are spaced apart to align with the state 
transition diagram on the right.
The diagram's node labels $\ket{xy}$ show the memory contents of a 
machine executing the program.
The arrows show possible transitions, weighted by the probabilities 
by which those transitions happen.
Randomized operations like \lstinline{rand_bit()} make multiple 
outgoing transitions from each node and are said to be 
\emph{non-deterministic}.
Non-randomized operations like \lstinline{y := x} create one 
transition out of every node and are said to be \emph{deterministic}.

The nodes are called \emph{worlds} because they represent different 
potential states of the world: for example, the $\ket{00}$ on the 
second row of the diagram represents a world in which the first 
\lstinline{rand_bit()} returned \lstinline{0}, while the $\ket{10}$ 
on its right represents an alternate reality in which the 
\lstinline{rand_bit()} gave \lstinline{1}.
A path in the transition diagram is called a \emph{world line}.
The probability that a world line is realized (i.e.\ that execution 
follows that path) is given by the product of the weights on the 
arrows along that path.

After executing each line in the program, the program can be in one 
of several worlds.
This uncertainty is captured by a \emph{superposition}, written as a 
sum of worlds weighted by their likelihoods.
For instance, the program state 
$\frac{1}{2}\ket{00} + \frac{1}{2}\ket{10}$ means $x,y = 0,0$ or 
$x,y = 1,0$ with equal probability at that point in the program.
In \cref{fig:eg:cpp}, the transition diagram's nodes are arranged 
into rows of such sums.
These sums are just a notation for vectors over $\mathbb{R}$, called 
\emph{state vectors}, which list the weights of all worlds.
For example, $\frac{1}{2}\ket{00} + \frac{1}{2}\ket{10}$ denotes the 
vector $[\frac{1}{2},0,\frac{1}{2},0]^T$, where we listed 
coefficients for $\ket{00}, \ket{01}, \ket{10}, \ket{11}$ in that 
order.
The operations in the language denote mappings between state vectors.
A single-world state vector like $\ket{xy}$ is called a \emph{basis 
  state}.
In matrix notation, a basis state has 1 in exactly one place, with 
0 everywhere else.

Not all vectors and mappings between them are implementable, however.
For the weights to make sense as probabilities, we need two 
commonsense properties:
\begin{itemize}
\item Law of total probability: probabilities appearing in a program 
  state must sum to 1, and operations must preserve this total.
\item Linearity: the state vector after any operation should be a 
  linear function of the state vector before the operation.
  Equivalently, the probability of landing in a given world should 
  equal the sum of the probabilities of realization of all world 
  lines leading up to that world.
\end{itemize}
Linearity means that operations form matrices.
By convention, its $j$-th \emph{column} lists the probabilities of 
transitions going out of the $j$-th basis sate.

\section{Quantum Programming as Probabilistic Programming}
\label{sec:qppl}

In this section, we derive quantum programming as a modification to 
classical probabilistic programming by allowing for negative 
transition weights, turning CPPL into a quantum programming language 
QPPL.
We explain how this modification leads to such concepts as 
destructive interference, unitarity, and reversibility, as well as 
the changes it mandates on programming languages.

In QPPL, a state vector $\sum_x q_x \ket{x}$ has real but possibly 
negative weights $q_x \in [-1..1]$ with $\sum_x |q_x|^2 = 1$, i.e.\ 
it has vector length 1 instead of total 1.
Each $q_x$ is called a \emph{probability amplitude}, or 
\emph{amplitude} for short.
Amplitudes are related to (classical) probabilities as follows: like 
in CPPL, a QPPL program ending in state vector $\sum_x q_x \ket{x}$ 
produces output by choosing a world randomly from that sum and having 
that world dictate the output, but in QPPL, world $\ket{x}$ is chosen 
with probability $|q_x|^2$.
Thus, amplitudes still measure likelihoods, just on a different scale 
than probabilities, and with sign.

The addition of sign with the square relationship with probabilities 
is the \emph{only} modification we impose on probabilistic 
programming in this section; all other changes fall out as 
byproducts.
For instance, the vector length being 1 is just the law of total 
probability, accounting for the square relationship.
Operations must preserve these lengths instead of totals.
Operations remain linear, though note that they are linear in $q_x$, 
not $|q_x|^2$.

\begin{remark}
  The reader might find the square relationship between amplitudes 
  and probabilities rather arbitrary, but it's more or less 
  mathematically forced.
  Aaronson \cite{Aaronson2004QuantumMechanicsIsland} gives an 
  elementary proof that other exponents lead either to classical 
  probabilistic programming or to a trivial programming model where 
  all operations are deterministic.
  Gleason's 
  theorem 
  gives an even more general argument, though its assumptions need 
  physical intuitions to justify.
\end{remark}

\begin{lrbox}{\myifbox}
\begin{lstlisting}
if x == 1:
  qnegate()
\end{lstlisting}
\end{lrbox}
\begin{figure}[t]
  \centering
  \[
    \begin{tikzcd}[row sep=1ex,/tikz/column 1/.append style={anchor=base west}]
      \lst{def main(x,y : bit):} &&
      & \ket{00}\ar[ddl,"\sqrthalf"']\ar[ddr,"\sqrthalf"]
      \\
      \hspace{2em}\lst{qrand_bit(x)}
      \\
      &&
      \sqrthalf\ket{00} \ar[dd,"1"']
      & +
      & \sqrthalf\ket{10} \ar[dd,"{\color{red}{-1}}"]
      \\
      \hspace{2em}\usebox{\myifbox}
      \\
      &&
      \sqrthalf\ket{00} \ar[dd,"\sqrthalf"'] \ar[ddrr,"\sqrthalf" pos=5/6]
      & +
      & \bigl({\color{red}{-\sqrthalf}}\bigr)\ket{10} \ar[ddll,"\sqrthalf"' pos=5/6] \ar[dd,"{\color{red}{-\sqrthalf}}"]
      \\
      \hspace{2em}\lst{qrand_bit(x)}
      \\
      &&
      {\color{red}{0}}\ket{00} & + & \ket{10} \ar[dd,"1"]
      \\
      \hspace{2em}\lst{y ^= x}
      \\
      \hspace{2em}\lst{return x,y}
      &&
      & & \ket{11}
    \end{tikzcd}
  \]
  \caption{Example of a quantum probabilistic program and its transition diagram.}
  \label{fig:eg:qpp}
\end{figure}

\Cref{fig:eg:qpp} illustrates the resulting changes to programming.
The quantum coin-toss \lstinline{qrand_bit(x)} takes a variable as 
argument, and its effect depends on the variable's value: if 
\lstinline{x} is 0, then it sets \lstinline{x} to 0 or 1 with equal 
probability amplitudes; if \lstinline{x} is 1, then it sets 
\lstinline{x} to 0 or 1 with equal but opposite-sign 
amplitudes.\footnote{Conventional quantum programming calls this the 
  Hadamard gate.}
Whereas splitting into two equal probabilities meant splitting into 
$\frac{1}{2}$ and $\frac{1}{2}$, with amplitudes it means splitting 
into $\sqrthalf$ and $\sqrthalf$, due to the scale difference.
Instead of assignment \lstinline{y := x}, QPPL uses XOR-assignment 
\lstinline{y ^= x}, which sets \lstinline{y} to the bitwise XOR of 
\lstinline{y} and \lstinline{x}.
This is because of reversibility, discussed below.

An \lstinline{if} statement works just like in classical 
probabilistic programming: each world independently evaluates the 
condition to decide whether or not to execute the body.
In this case, the body \lstinline{qnegate()} does nothing with 
probability amplitude $-1$, i.e.\ with certainty but with negative 
amplitude.
The effect is to flip the sign of the probability amplitude of being 
in the current world.
Put together, \lstinline{if x == 1: qnegate()} flips the amplitude's 
sign in those worlds where \lstinline{x} is 
\lstinline{1}.\footnote{Conventional quantum 
  programming calls this the Z gate.}

The main difference made by negative weights is the possibility of 
cancellation, known as \emph{destructive interference}.
As before, we land in a world with probability amplitude given by the 
sum of products of amplitudes on all incoming world lines.
But since amplitudes can be negative, they can sum to zero, like in 
the bottom-left node of the diagram in \cref{fig:eg:qpp}.
Thus, after the second call to \lstinline{qrand_bit(x)}, the example 
program has \emph{zero} chances of landing in world $\ket{00}$.

Destructive interference causes a subtle change in the relationship 
between the worlds.
Before, the worlds were completely independent trials that never 
interact, as only one of them could really happen at a time.
With destructive interference, we must think that all worlds are 
simultaneously real, interacting to decide which possibilities live 
and which ones die.
But this interaction happens only when worlds evolve to identical 
states, which is tricky to control under reversibility constraints 
discussed below.
Using it to consolidate information from multiple worlds is the 
central challenge in quantum programming 
\cite{Aaronson2021WhatMakesQuantum}.

The law of total probability on amplitudes has two interrelated 
consequences.
\begin{itemize}
\item Preservation of orthogonality: because the dot product can be 
  expressed by vector lengths,\footnote{Known as the polarization 
    identity: for real vectors, 
    $u \cdot v = \frac{1}{2}(\veclen{u + v}^2 - \veclen{u}^2 - 
    \veclen{v}^2)$.} preserving lengths means preserving dot 
  products, hence preserving orthogonality.
  Basis states are orthogonal, so all operations must map them to 
  orthogonal states, i.e.\ the corresponding matrix's columns must be 
  orthogonal.
  This property is called \emph{isometricity} or, in the absence of 
  memory allocation, \emph{unitarity}.
\item Reversibility: as linear maps are non-injective precisely when 
  they collapse several dimensions into one (or to the trivial space 
  $\{0\}$), every orthogonality-preserving matrix is injective.
  Thus, a quantum program must be built up from only injective 
  operations.
  Such a program is said to be \emph{reversible}, as it can be 
  executed in reverse, undoing its operations one by one 
  \cite{GluckYokoyama2023ReversibleComputingProgramming}.
\end{itemize}
These are prices to be paid for negative amplitudes and destructive 
interference.

Preservation of orthogonality explains the design of 
\lstinline{qrand_bit(x)}, characterized as the simplest 
non-deterministic operation, an analogue of the classical coin toss.
To have orthogonal (hence disinct) columns, it must change its 
behavior depending on the world, i.e.\ on preexisting memory 
contents.
Thus, it must read a variable \lstinline{x} before deciding how to 
generate the random bit.
As the simplest non-deterministic operation, it shouldn't read any 
other variables if possible; but that forces it to write to 
\lstinline{x} as well, for the only alternative is to write to some 
\lstinline{y} without reading it, failing to behave differently on 
worlds differing only in the preexisting value of \lstinline{y}.
If we want the operation to give \lstinline{0} and \lstinline{1} with 
equal amplitudes (as a coin toss should) at least when the 
preexisting value of \lstinline{x} is \lstinline{0}, the transition 
matrix (in the simple case where \lstinline{x} is the only program 
variable) must have the form 
$\frac{1}{\sqrt{2}}\begin{bmatrix}1 & a \\ 1 & b\end{bmatrix}$.
Column orthogonality and normalization force $a = -b = \pm1$, which 
gives \lstinline{qrand_bit(x)} up to a choice of sign.

Reversibility induces a more pervasive change, which has been studied 
extensively in classical languages 
\cite{GluckYokoyama2023ReversibleComputingProgramming}.
An assignment like \lstinline{y := x} is quintessentially 
irreversible, as it erases information about the value previously 
held by \lstinline{y}.
A standard trick is to use XOR-assignment \lstinline{y ^= x} instead, 
which can be undone by executing \lstinline{y ^= x} once more.
But for this reversal to work, the value of \lstinline{x} must remain 
undisturbed by the modification to \lstinline{y}.
QPPL ensures this by requiring that the right-hand side \lstinline{x} 
(which in general can be an expression referencing multiple 
variables) does not mention the assigned-to variable \lstinline{y}.

A similar provision is needed for conditionals.
To undo a conditional like \lstinline{if COND: BODY}, it is necessary 
to know whether the \lstinline{BODY} was executed and therefore needs 
to be undone.
For this, it is necessary to preserve the value of \lstinline{COND} 
across the execution of \lstinline{BODY}.
QPPL ensures this by requiring that \lstinline{BODY} does not assign 
to variables mentioned in \lstinline{COND}.

Note that QPPL requires reversibility only at the level of 
statements---things that can appear on lines of their own, like 
XOR-assignments, \lstinline{if} blocks, \lstinline{qrand_bit(x)}, or 
\lstinline{qnegate()}.
The parts of statements that calculate values (called expressions), 
like \lstinline{COND} above or the right-hand side of an 
XOR-assignment, need not be reversible.
They can use non-injective functions like logical AND or arithmetic, 
but in exchange, they cannot contain statements like XOR-assignment.

\begin{remark}
  Readers familiar with conventional, qubit-based quantum programming 
  may have noticed that QPP is really ``just'' a change in 
  perspective:
  \begin{itemize}
  \item QPP views a computation from an observer inside the computer, 
    running along a particular world-line and describing how each 
    step evolves the current world into one or more futures.
  \item Conventional quantum programming views the computation from 
    an observer external to the worlds who can see all the branching 
    structure.
  \end{itemize}
  For readers familiar with the list monad, the former is analogous 
  to coding in monadic style in that monad, while the latter is like 
  direct manipulation of lists.
  Quantum non-determinism is not quite a monad but has return and 
  bind that satisfy the monad laws 
  \cite{MuBird2001FunctionalQuantumProgramming,%
    VizzottoEtAl2006StructuringQuantumEffects}.
  The point of this paper is to show what a conceptual difference 
  this change in perspective makes.
\end{remark}

\begin{remark}\label{rmrk:complex-numbers}
  Quantum theory famously requires complex numbers to describe, so 
  the reader may wonder why we only mention real numbers, and whether 
  that constitutes a limitation.
  QPP can in fact be formulated equally well with complex numbers.
  Just change ``negative probabilities'' to ``complex 
  probabilities'', $q_x \in [0..1]$ to 
  $q_x \in \mathbb{C} \land |q_x| \leq 1$, and ``square'' to ``square-absolute 
  value'', and most of this paper stands, verbatim.
  We nonetheless left out complex numbers because:
  \begin{itemize}
  \item They are not necessary.
    A complex-amplitude state vector $\sum_x (a_x + ib_x)\ket{x}$ with 
    $a_x,b_x \in \mathbb{R}$ can be encoded by real amplitudes as 
    $\sum_x a_x\ket{x0} + \sum_x b_x\ket{x1}$ with just 1 bit of 
    overhead.
    This simple encoding can implement any complex-amplitude 
    computation with only constant slowdown 
    \cite{BernsteinVazirani1993QuantumComplexityTheory,%
      Aharonov2003SimpleProofToffoli}.
  \item They are less intuitive.
    What they add is the ability to cancel along axes other than the 
    real line.
    While this extra freedom is handy in algorithms like Shor's, 
    introducing complex linear algebra with its lesser known inner 
    product seems ill-motivated until one gets to such algorithms.
    Complex or real, in QPP it is the ability to have outcomes occur 
    with opposite signs and cancel that defines quantum programming, 
    and we feel the extra freedom obscures this idea.
  \end{itemize}
  Whether generalizing to complex numbers is worth the conceptual 
  complication is an interesting open question.
  Because complex amplitudes can be so simply encoded by real ones, 
  we suspect one can comfortably program all the way with real 
  numbers, but whether that works in practice remains to be seen.
\end{remark}

\section{Programming Example}
\label{sec:examples}

In this section, we show an example that illustrates what it is like 
to program in QPPL.
This shows how the QPP point of view can be a useful device for 
explaining or motivating quantum algorithms, at least in simple 
cases.

\cref{fig:deutsch} shows Deutsch's algorithm in QPPL.
The task it performs is: given a black-box mapping \lstinline{f} from 
a single bit to a single bit, determine if \lstinline{f} is a 
constant mapping (returning 1 if non-constant).
The displayed algorithm does so in time taken to call \lstinline{f} 
only once (with constant overhead), no matter how long \lstinline{f} 
takes, which is evidently impossible in classical computing.
For simplicity, \lstinline{f} is treated as a macro that can be used 
in the condition expression of an \lstinline{if} statement.
There are several techniques to represent it with a user-defined 
function instead (such as injectivization 
\cite{GluckYokoyama2023ReversibleComputingProgramming} or 
irreversible functions \cite{Omer2003StructuredQuantumProgramming}), 
but user-defined functions are out of scope for this paper.

The crux of the algorithm is the dotted box in \cref{fig:deutsch}.
Prior to that box, we set up two worlds, one evaluating 
\lstinline{f(0)} and the other evaluating \lstinline{f(1)}.
We do this because judging constancy of \lstinline{f} requires 
checking the values of both \lstinline{f(0)} and \lstinline{f(1)}, 
but to do that in time taken for one call to \lstinline{f}, we have 
no choice but to split the work between different worlds.
Now, in classical probabilistic programming, we'd be stuck, for there 
is no way to consolidate information from these worlds, which are 
independent trials and therefore never interact.
However, in quantum programming, worlds can interact via destructive 
interference.
Conversely, destructive interference is our only hope of performing 
this classically impossible feat, for as explained in 
\cref{sec:qppl}, destructive interference is the only essential 
difference between quantum and classical computing.

\begin{figure}[t]
  \centering
    \begin{tabular}{ll}
\begin{lstlisting}
f : bit -> bit

def main(x : bit):
  qrand_bit(x)
  if f(x) == 1:
    qnegate()
  qrand_bit(x)
  return x
\end{lstlisting}
      &
    $
        \begin{tikzcd}[column sep=small,
          /tikz/execute at end picture={
            \node (box) [rectangle, draw, dotted, fit=(TL) (BL) (BR)] {};
          }]
        & \ket{0} \dlar["\sqrthalf"'] \drar["\sqrthalf"] \\
        \sqrthalf\ket{0} \dar["{\color{red}{(-1)^{f(0)}}}"']
        & + &
        \sqrthalf\ket{1} \dar["{\color{red}{(-1)^{f(1)}}}"]
        \\
        |[alias=TL]| {\color{red}{\frac{(-1)^{f(0)}}{\sqrt{2}}}}\ket{0} \dar["\sqrthalf"'] \ar[drr,"\sqrthalf" pos=6/7]
        & + &
        |[alias=TR]|{\color{red}{\frac{(-1)^{f(1)}}{\sqrt{2}}}}\ket{1} \dar["{\color{red}{-\sqrthalf}}"'] \ar[dll,"\sqrthalf"' pos=6/7]
        \\
        |[alias=BL]| {\color{red}{\frac{(-1)^{f(0)} + (-1)^{f(1)}}{2}}}\ket{0}
        & + &
        |[alias=BR]| {\color{red}{\frac{(-1)^{f(0)} - (-1)^{f(1)}}{2}}}\ket{1}
      \end{tikzcd}
      $
    \end{tabular}
  \caption{Deutsch's algorithm in QPPL.
    Source lines are not aligned with the transition diagram to 
    conserve space.}
  \label{fig:deutsch}
\end{figure}

For two worlds to destructively interfere, we need to steer them 
towards identical states, or memory contents.
The two worlds so far have memory $\ket{0}$ and $\ket{1}$, so they 
meet by having either the 0 rewritten to 1 or vice versa.
Which should it be?
Well, both: reversible, deterministic operations can only permute 
worlds, so worlds can only meet during a non-deterministic operation 
like \lstinline{qrand_bit(x)}, which performs both rewrites.

As seen from the arrows in the dotted box, \lstinline{qrand_bit(x)} 
makes two equal-amplitude transitions into $\ket{0}$ and two 
opposite-signed transitions into $\ket{1}$.
Thus, weights from the two worlds add on the $\ket{0}$ side whereas 
they subtract on the $\ket{1}$ side.
Hence, to get cancellation on the desired side, it suffices to steer 
the worlds' weights before \lstinline{qrand_bit(x)} to be equal when 
$f(0) = f(1)$, but equal and opposite-signed when $f(0) \neq f(1)$.
The program achieves that by calling \lstinline{qnegate()} in just 
those worlds with \lstinline{f(x) == 1}.

Motivating the algorithm through QPP like this is not only useful for 
understanding the algorithm but also gives lessons on the benefits 
and challenges of quantum computing in general.
Namely, negative amplitudes let us consolidate information from many 
worlds, but reversibility makes that tricky, which is why experts 
warn against na\"ively thinking of the worlds as parallel processors 
\cite{Aaronson2021WhatMakesQuantum}.
It remains to be seen if QPP aids in developing general intuitions 
that help construct more advanced algorithms.


\section{Measurements}
\label{sec:measurements}

In this section, we add a layer of classical probability on top of 
the model from \cref{sec:qppl}, which gives us a \emph{two-layer 
  probability model}.
This model lets us explain measurements, along with why it is 
involved in storage deallocation or output.

Most existing quantum programming languages have a 
\lstinline{measure} operation, whose behavior is explained as 
irreversibly collapsing qubits into classical states.
Concretely, it maps a qubit in superposition 
$q_0\ket{0} + q_1\ket{1}$ to a state with definite classical value, 
either $\ket{0}$ or $\ket{1}$ with classical probabilities $|q_0|^2$ 
and $|q_1|^2$, respectively.
Measurement can also be done on a subset of qubits in the program.

Measurement also happens implicitly at the end of every program, when 
we read the program's output.
The output is determined by collapsing the final superposition to a 
single world, just like how a classical probabilistic program 
executes by deciding on a single outcome.
Measurement also happens implicitly when discarding data: 
reversibility requires that no data is ever thrown away, so 
measurement is needed to convert it to classical data first.

How can we understand measurements in QPP, which has no notion of 
qubits?
In QPP, output, discarding, and measurement can all be understood 
uniformly as losses of chance for destructive interference.
We've mentioned that interference happens when multiple worlds 
converge on the same program state, but to be completely accurate, 
they must converge on the same state for everything in the entire 
universe.
We can get away with focusing exclusively on the machine state 
because a quantum computer is sufficiently isolated from its 
surroundings that its state is decoupled from the rest of the 
universe.
Output and discarding break this isolation.

Ink on paper, electrical state of classical RAM, memory stored in a 
human brain---whatever the form, output affects a part of ``the rest 
of the universe'' in a way that depends on parts of program memory.
If a world wants to interfere with another world that had different 
data in those parts of memory, it not only has to merge the memory 
contents but also hunt down and modify all traces of the output so 
that they match as well, an impossibility.
Likewise, deallocation is a vow not to touch a part of the storage 
anymore, effectively turning it into a part of ``the rest of the 
universe'' that cannot be brought to cooperate with interference.
Explicit measurement is similar: conceptually, \lstinline{measure} is 
an operation that reads out the values of the measured bits into a 
classical medium, where ``classical'' means ``does not cooperate with 
interference''.

Mathematically, loss of interference can be modeled as conversion of 
amplitudes to probabilities.
Recall that in QPP, negative amplitudes and the destructive 
interference they enable are \emph{the} defining feature of quantum.
Taking for simplicity the case where all bits in the program are 
measured, all future destructive interference is eliminated, so the 
superposition resulting at that point may as well be classical.
In effect, the measurement has converted amplitudes to probabilities.
To capture this idea, we enrich program states from state vectors to 
classical probability distributions over them.\footnote{These 
  distributions are known as \emph{ensembles} in quantum theory 
  \cite[Chapter~2]{NielsenChuang2011QuantumComputationQuantum}.}

A \emph{two-layer state} is a list of pairs 
$[(p_j, \sum_{x \in 2^n} q_{j,x} \ket{x})]_{j=1}^m$ where:
\begin{itemize}
\item The $p_j$'s are classical probabilities, i.e.\ real numbers in 
  $[0..1]$ that add to 1.
\item For each $j$, the $\sum_{x \in 2^n} q_{j,x}\ket{x}$ is a valid 
  (i.e.\ length-1) state vector called the $j$-th \emph{branch}.
  World $\ket{x}$ in the $j$-th branch is called the $j,x$-th 
  \emph{world} of the two-layer state.
\end{itemize}
If a program is in two-layer state 
$[(p_j, \sum_x q_{j,x} \ket{x})]_j$, then the machine is actually in a 
single branch, the $j$-th one with probability $p_j$.
The $p$'s thus reflect our ignorance of which branch the machine is 
in.
If the machine is in the $j$-th branch, then it's also in the 
$j,x$-th world with probability amplitude $q_{j,x}$, which can 
destructively interfere with other worlds in the $j$-th branch.
The $q$'s reflect the quantum non-determinism explained in 
\cref{sec:qppl}.

Measuring the $m$ rightmost bits in an $n$-bit program ($m \leq n$) 
evolves the two-layer state by the following function:
\begin{equation}\label{eq:measure}
  \meas{n}{m} \biggl[\biggl(p_j, \sum_{x \in 2^{n-m},y \in 2^m}q_{j,x,y}\ket{xy}\biggr)\biggr]_j = \biggl[\biggl(p_jQ_{j,y}^2, \sum_{x} \frac{q_{j,x,y}}{Q_{j,y}} 
  \ket{xy}\biggr)\biggr]_{j,y}
\end{equation}
where $Q_{j,y} := \sqrt{\sum_x |q_{j,x,y}|^2}$.
In each branch, the worlds with a common $y$ value are collected into 
a new branch, moving the $y$ index from the $\sum$ to the brackets 
marking the list of branches.
The probability amplitude of being in those worlds (the $Q_{j,y}$) 
gets squared and turned into probability.
Effectively, the $q$'s capture quantum non-determinism introduced by 
\lstinline{qrand_bit}, which can produce destructive interference, 
and the $p$'s are the mass of amplitudes that has lost that 
capability to measurement.

\begin{example}
  \cref{fig:qec} shows a simplified example of quantum error 
  correction.
  A logical bit is triplicated into physical bits \lstinline{x}, 
  \lstinline{y}, \lstinline{z}.
  An unwanted rogue particle that enters the quantum computer and 
  causes an error is modeled as a bit \lstinline{r}.
  (Though the particle is not a part of the device's memory, we can 
  model it as if it is.)
  The error caused by the rogue particle is a bit-flip error in 
  \lstinline{z}, whose effect is modeled by \lstinline{z ^= r}.
  The error correction code tests for mismatches among \lstinline{x}, 
  \lstinline{y}, \lstinline{z} using XOR (\lstinline{^}), stores the 
  results to ancillary (i.e.\ temporary) storage \lstinline{a}, 
  \lstinline{b}, and measures them.
  The rogue particle also exits the device around this time, causing 
  an information leak also modeled as measurement.
  The sequence of \lstinline{if}'s then corrects for the error, 
  making \lstinline{x}, \lstinline{y}, \lstinline{z} equal again.

  Let us trace this code in the two-layer probability model.
  Assume \lstinline{x}, \lstinline{y}, \lstinline{z} are initialized 
  to $q_0\ket{000} + q_1\ket{111}$, representing the logical bit 
  $q_0\ket{0} + q_1\ket{1}$, while the rogue particle starts out in 
  state $r_0\ket{0} + r_1\ket{1}$ (where 
  $|q_0|^2 + |q_1|^2 = |r_0|^2 + |r_1|^2 = 1$).
  Ancillary bits \lstinline{a} and \lstinline{b} are initialized to 
  0.
  Coming from different sources, the rogue particle's value is 
  independent of the program variables \lstinline{x}, \lstinline{y}, 
  \lstinline{z}, \lstinline{a}, \lstinline{b}, so just like in 
  classical probability theory, the probability amplitude of 
  $xyzab=00000 \land r=0$ is the product of the amplitudes of 
  $xyzab = 00000$ and $r = 0$.
  Thus, writing worlds in the format $\ket{xyz,abr}$, the probability 
  amplitude of being in world $\ket{000,000}$ is $q_0r_0$.
  Analyzing all combinations of values for $xyz,abr$ in the same 
  way,\footnote{This is known in conventional quantum programming as 
    taking the \emph{tensor product}.}
  we get the first two-layer state shown in the figure.

  All operations before \lstinline{measure} are deterministic, so 
  they simply act independently on each world, giving the second 
  two-layer state in the figure.
  We want to restore triplication, getting a state of the form 
  $q_0\ket{000,abr} + q_1\ket{111,abr}$, but no reversible operation 
  can do that: $abr$ differs across worlds, carrying information 
  about which world saw what kind of error, and eliminating the error 
  requires erasing this information, which reversible operations 
  cannot do.

  Measuring \lstinline{a}, \lstinline{b}, \lstinline{r} splits the 
  state into branches according to those variables' values.
  Two worlds have $abr = 000$, with amplitudes $q_0r_0$ and $q_1r_0$, 
  so the likelihood of transitioning to a branch with $abr=000$ is 
  $|q_0r_0|^2 + |q_1r_0|^2 = |r_0|^2$.
  Those worlds thus form a branch realized by probability $|r_0|^2$, 
  with amplitudes rescaled so that the branch has vector length 1.
  Worlds with $abr=011$ are treated likewise, giving the third 
  two-layer state in the figure.

  Then the \lstinline{if}'s correct the bit-flip error.
  At a high level, this is made possible because \emph{in each 
    branch}, the value of $abr$ is now uniform across worlds and no 
  longer carries information that needs to be erased.
  The $abr$ now identifies the branch instead, so the information it 
  contains has effectively moved from the amplitude to the classical 
  layer.
\end{example}

\newsavebox{\initstate}
\savebox{\initstate}{\ensuremath{\left[\left(1,
        \left(\begin{array}[c]{l}
          q_0r_0 \ket{000,000} + q_1r_0\ket{111,000} \\
          \hspace{1em}+ q_0r_1\ket{000,001} + q_1r_1\ket{111,001}
        \end{array}\right)\right)\right]}}

\newsavebox{\premeasure}
\savebox{\premeasure}{\ensuremath{\left[\left(1,
        \left(\begin{array}[c]{l}
          q_0r_0\ket{000,000} + q_1r_0\ket{111,000} \\
          \hspace{1em}+ q_0r_1\ket{001,011} + q_1r_1\ket{110,011}
        \end{array}\right)\right)\right]}}

\newsavebox{\postmeasure}
\savebox{\postmeasure}{\ensuremath{\left[
            \begin{array}{l}
              \left(|r_0|^2, q_0\ket{000,000} + q_1\ket{111,000}\right) \\
              \hspace{0.1em},\hspace{0.1em}
              \left(|r_1|^2, q_0\ket{001,011} + q_1\ket{110,011}\right)
            \end{array}
          \right]}}

\newsavebox{\postcorrect}
\savebox{\postcorrect}{\ensuremath{\left[
            \begin{array}{l}
              \left(|r_0|^2, q_0\ket{000,000} + q_1\ket{111,000}\right) \\
              \hspace{0.1em},\hspace{0.1em}
              \left(|r_1|^2, q_0\ket{000,011} + q_1\ket{111,011}\right)
            \end{array}
          \right]}}

\begin{figure}[t]
  \begin{center}
    \[
      \begin{tikzcd}[row sep=0.2ex,column sep=1em,/tikz/column 1/.append style={anchor=base west},
        execute at end picture={%
          \node[label={\usebox\initstate}, yshift=-1.5\baselineskip]
          at (\tikzcdmatrixname-2-2) {};
          \draw (\tikzcdmatrixname-1-1.south east) -- (\tikzcdmatrixname-2-2.west);
          \node[label={\usebox\premeasure}, yshift=-1.5\baselineskip]
          at (\tikzcdmatrixname-4-2) {};
          \draw (\tikzcdmatrixname-4-1.south east) -- (\tikzcdmatrixname-4-2.west);
          \node[label={\usebox\postmeasure}, yshift=-1.5\baselineskip]
          at (\tikzcdmatrixname-6-2) {};
          \draw (\tikzcdmatrixname-5-1.south east) -- (\tikzcdmatrixname-6-2.west);
          \node[label={\usebox\postcorrect}, yshift=-1.5\baselineskip]
          at (\tikzcdmatrixname-8-2) {};
          \draw (\tikzcdmatrixname-8-1.south east) -- (\tikzcdmatrixname-8-2.west);
        }]
        \texttt{... \# init x,y,z,a,b,r}
        & {\hspace{\wd\initstate}}
        \\
        {\lst{z ^= r}\texttt{ \# bit-flip error}}
        & {\hspace{\wd\initstate}}
        \\
        \lst{a ^= (x ^ y)}
        & {\hspace{\wd\initstate}}
        \\
        \lst{b ^= (y ^ z)}
        & {\hspace{\wd\premeasure}}
        \\
        \lst{measure (a,b,r)}
        & {\hspace{\wd\premeasure}}
        \\
        \lst{if a,b == 0,1: z ^= 1}
        & {\hspace{\wd\postcorrect}}
        \\
        \lst{if a,b == 1,0: x ^= 1}
        & {\hspace{\wd\postcorrect}}
        \\
        \lst{if a,b == 1,1: y ^= 1}
        & {\hspace{\wd\postcorrect}}
      \end{tikzcd}
    \]
  \end{center}
  \caption{A model of quantum error correction, omitting 
    \lstinline{def main} and initialization.
    On the right are the intermediate two-layer states at select 
    points in the code indicated by line segments.
    Worlds are written in $\ket{xyz,abr}$ format.}
  \label{fig:qec}
\end{figure}

QPP's understanding of measurement also suggests how we can avoid the 
crippling effects of information leakage.
The trouble was that leaked values preserve differences in the 
worlds, preventing destructive interference.
It follows that if the leaked bits happen to have the same value in 
all worlds, then no interference is lost.
This leads to the idea of \emph{uncomputation}: setting temporary 
bits to the same value (usually 0) in all worlds (in each branch) 
before discarding them.


QPPL allows storage allocation by a \lstinline{new} statement:
\begin{center}
  \begin{tabular}{c}
\begin{lstlisting}[mathescape=true]
def main (x : bit):
  new y := $\neg$x # y initialized to complement of x
  new z := y
  return y # x,z are discarded
\end{lstlisting}
  \end{tabular}
\end{center}
Though user-defined functions are beyond the scope of this paper, if 
they are added, then it is most natural that storage allocated inside 
them are discarded when the function returns, except for those 
returned as part of the function's return value.
In that case, it is the programmer's responsibility to ensure proper 
uncomputation if destructive interference is desired thereafter.

Measurement cannot be used inside conditionals because that's 
unimplementable: it would involve leaking information in some worlds 
and not others, requiring a measurement that cooperates with quantum 
non-determinism, an oxymoron.
QPPL also forbids conditional allocation (useful in dynamic 
allocation scenarios), though this is just for simplicity.

Automatic verification and generation of uncompute code have been 
studied extensively \cite{RandEtAl2019ReQWIREReasoningReversible,%
  BichselEtAl2020SilqHighlevelQuantum,%
  ParadisEtAl2021UnqompSynthesizingUncomputation}, and a programming 
language that deals with iteration and dynamic allocation is also 
known \cite{PechouxEtAl2020QuantumProgrammingInductive}.
Growing QPPL into a full-fledged programming language offering such 
conveniences is left for future work.

\section{Formal Semantics}
\label{sec:formal-semantics}

In this section, we give a formal semantics to a core subset of QPPL.
We prove that its operations are both implementable (i.e. definable 
by unitaries, measurements, and initialization of new qubits) and 
computationally universal (i.e. it can efficiently approximate all 
possible quantum programs, provided data is encoded in a certain 
way).
This section assumes familiarity with quantum computing and 
programming language theory.

\newcommand{\xorassign}{\ensuremath{\mathbin{\texttt{\string^=}}}}
\newcommand{\FV}{\operatorname{FV}}
\newcommand{\AV}{\operatorname{AV}}
\newcommand{\qneg}{\texttt{qneg}\,()}
\newcommand{\qrand}[1]{\texttt{qrand}\,(#1)}
\newcommand{\measure}[1]{\texttt{measure}\,(#1)}
\newcommand{\new}[1]{\texttt{new}\,(#1)}
\newcommand{\seq}[2][{}]{\overline{#2}^{#1}}
\begin{figure}[t]
  \centering
  \begin{align*}
    x,y \in \textit{Vars} \\
    P \in \textit{Prog} &::= \texttt{def main}\,(\seq{x})(\seq{S}) \\
    S \in \textit{Stmt} &::= \new{\seq{x}} \mid \measure{\seq{x}} \mid C \\
    C \in \textit{Comp} &::= \texttt{if}\,E\,(\seq{C})\ [\FV(E) \cap \AV(\seq{C}) = \varnothing] \mid x \xorassign E\ [x \notin \FV(E)] \\
                       &\phantom{::=}\mid \qrand{x} \mid \qneg \\
    E \in \textit{Expr} & ::= x \mid 0 \mid 1 \mid \neg E \mid E \land E \mid E \lor E
  \end{align*}
  \begin{mathpar}
    \AV(C,\seq{C'}) = \AV(C) \cup \AV(\seq{C'}) \and 
    \AV(\texttt{if}\,E\,(\seq{C})) = \AV(\seq{C}) \and 
    \AV(x \xorassign E) = \{x\} \and 
    \AV(\qrand{x}) = \{x\} \and 
    \AV(\qneg) = \varnothing
  \end{mathpar}
  \caption{Syntax of core QPPL.  Production rules marked $[\phi]$
    apply only if $\phi$ is true; for example, $x \xorassign E$ is a
    valid expression only if $x \notin \FV(E)$.}
  \label{fig:syntax}
\end{figure}

The syntax of core QPPL is formalized in \cref{fig:syntax}.
We write sequences of length $n$ like $\seq[n]{S}$ or simply 
$\seq{S}$ if the length is unimportant or understood from context.
\textit{Comp} is the subset of statements usable inside conditionals, 
called \emph{computational statements}.
$\FV(E)$ stands for the free variables in an expression $E$ 
(definition omitted).
$\AV(\overline{C})$ stands for the variables assigned to in $\overline{C}$.

The syntax is mostly as presented before, with the following 
simplifications:
\begin{itemize}
\item All variables are implicitly of type \lstinline{bit}, and 
  expressions are built from standard Boolean operators.
\item \lstinline{new} omits initialization expressions.
  All variables are initialized to zero, and the programmer can set 
  them to any desired value with XOR-assignment.
\item There is no explicit \lstinline{return} statement.
  All live variables are returned implicitly as quantum data at the 
  end of the program.
  An alternative semantics where all bits are implicitly measured at 
  the end can be simulated by explicitly calling \lstinline{measure}.
\item Some names are slightly abbreviated (e.g.\ 
  \lstinline{qrand_bit} $\rightarrow$ \texttt{qrand}).
\end{itemize}

The reference semantics for QPPL is defined in terms of the two-layer 
probability model, as that model is an integral part of how semantics 
is explained to the programmer.
We will later show that the two-layer model translates to a more 
standard density matrix semantics.

A two-layer state is a probability distribution on an amplitude 
distribution.
An \emph{amplitude distribution} is like a probability distribution, 
except it assigns probability amplitudes to outcomes, whose squares 
add to 1.
Continuing from preceding sections, we write $\sum_x q_x \ket{x}$ for 
the amplitude distribution assigning amplitude $q_x$ to outcome 
$\ket{x}$, while writing $[(p_j, a_j)]_j$ for the probability 
distribution assigning probability $p_j$ to outcome $a_j$.
The two-layer state notation from before is a combination of these 
notations.

\newcommand{\Set}{\mathbf{Set}}
\newcommand{\FinSet}{\mathbf{FinSet}}

\begin{definition}
  For any finite set $X$, let $PX$ be the set of all probability 
  distributions over $X$.
  For functions $f : X \rightarrow Y$, define the linear map 
  $Pf : PX \rightarrow PY$ by $Pf[(1,x)] := [(1,f(x))]$.
  Similarly, let $QX$ be the set of all amplitude distributions and 
  set $Qf\ket{x} := \ket{f(x)}$.\footnote{In other words, $P,Q$ are 
    probability/amplitude distribution functors 
    $\FinSet \rightarrow \Set$.}
\end{definition}

The reference two-layer probability semantics is summarized in 
\cref{fig:two-layer-semantics}, using the following auxiliary 
definitions.
\begin{itemize}
\item $I_n$ is the $n\times n$ identity matrix and $I_m^n$ with 
  $m \leq n$ is $I_n$ cut short to $m$ columns (i.e.\ an inclusion 
  map $\mathbb{R}^m \rightarrow \mathbb{R}^n$).
\item $H$ is the $2\times 2$ Hadamard matrix.
\item $\operatorname{if}(e,c,a)$ is $c$ if $e = 1$ and $a$ otherwise.
\item $\meas{n}{m}$ is as defined in \cref{eq:measure} of 
  \cref{sec:measurements}.
\item $[f(j,j',k,k')]_{k \in n, k' \in n'}^{j \in m, j' \in m'}$ denotes 
  the $mm'\times nn'$ matrix $A$ whose $(j,j'),(k,k')$-entry is 
  $f(j,j',k,k')$.
\end{itemize}
When a sequence of statements $\seq{S}$ that allocates $k$ new 
variables appears in a context that defines the variables 
$\seq[n]{x}$, its semantics is a function 
$\sem{\seq{x} \vdash \seq{S}} : P(Q(2^n)) \rightarrow P(Q(2^{n+k}))$.
The semantics of a program is the semantics of its statements with 
the inputs as the defined variables.

The semantics of each statement is as described in previous sections.
For the handling of \lstinline{if}, we note that by straightforward 
induction, $\sem{\seq{x} \vdash \overline{C}}$ is always of the form 
$PU$ with linear $U$.
Moreover, $P$ is faithful (i.e.\ injective on functions), so 
$PU = \sem{\seq{x},\seq{y} \vdash \overline{C}}$ uniquely defines $U$.
Intuitively, this extracts the semantics of $\seq{C}$ restricted to 
the amplitude layer.
The semantics of $\texttt{if}\,E(\overline{C})$ is then 
$\sem{\seq{C}}$ in worlds where $E$ is true, but the identity (i.e.\ 
no-op) in other worlds.

\begin{figure}[t]
  \centering
  \begin{align*}
    \sem{\texttt{def main}\,(\seq{x})(\seq{S})} &= \sem{\seq{x} \vdash \seq{S}} \\
    \sem{\seq[n]{x} \vdash \new{\seq[m]{y}}, \seq{S}} &= P(Q(I_{2^n}^{2^{n+m}}));\sem{\seq{x}, \seq{y} \vdash \seq{S}} \\
    \sem{\seq[n]{x},\seq[m]{y} \vdash \measure{\seq{y}}, \seq{S}} &= \meas{n+m}{m};\sem{\seq{x}, \seq{y} \vdash \seq{S}} \\
    \sem{\seq[n]{x},\seq[m]{y} \vdash \texttt{if}\,E\,(\seq{C}), \seq{S}} &= P\left(\left[\bracket{\xi'\eta'}{\operatorname{if}(e(\xi),U,I_{2^{n+m}})}{\xi\eta}\right]_{\xi \in 2^n, \eta \in 2^m}^{\xi' \in 2^n, \eta \in 2^m}\right);\sem{\seq{x}, \seq{y} \vdash \seq{S}} \\
                                                &\text{where\hspace{1.5em}}
                                                  \begin{array}[t]{l@{\hspace{1.5em}}l}
                                                    \FV(E) \subseteq \seq{x} &
                                                                        \AV(\overline{C}) \subseteq \seq{y} \\
                                                    e = \sem{\seq{x} \vdash E} &
                                                                            PU = \sem{\seq{x},\seq{y} \vdash \overline{C}}
                                                  \end{array}
    \\
    \sem{x,\seq[n]{y} \vdash x \xorassign E, \seq{S}} &= P\left(\left[\operatorname{if}(\xi' = \xi \oplus e(\eta),1,0)\right]_{\xi \in 2, \eta \in 2^n}^{\xi' \in 2, \eta \in 2^n}\right); \sem{x,\seq{y} \vdash \seq{S}} \\
                                                &\text{where\hspace{1.5em}}
                                                  \begin{array}[t]{l@{\hspace{1.5em}}l}
                                                    \FV(E) \subseteq \seq{y} &
                                                                        e = \sem{\seq{x} \vdash E}
                                                  \end{array}
    \\
    \sem{\seq[n]{x},y \vdash \qrand{y}, \seq{S}} &= P(I_n\otimes H);\sem{\seq{x}, y \vdash \seq{S}} \\
    \sem{\seq[n]{x} \vdash \qneg, \seq{S}} &= P(-I_n);\sem{\seq{x} \vdash \seq{S}} \\
    \sem{\seq[n]{x} \vdash E} &= \text{(the obvious interpretation as a function $2^n \rightarrow 2$)}
  \end{align*}
  \caption{Two-layer probability semantics for core QPPL.
    We implicitly coerce matrices to linear maps: e.g. $P(-I_n)$ is 
    $P$ applied to the negation map $v \mapsto -v$.
    Semicolons denote composition of maps.
    We omit the permutation matrices needed to reorder the variables 
    so that statements match the left-hand sides of these 
    definitions.}
  \label{fig:two-layer-semantics}
\end{figure}

\begin{lemma}
  \label{thm:unitarity}
  $\sem{\seq{x} \vdash \overline{C}}$ is unitary.
  \begin{proof}
    Induction with the inductive hypothesis strengthened with: if 
    $\seq{x}$ has length $n+m$, the last $m$ of which equals 
    $\AV(\overline{C})$, then $\bracket{\xi \eta}{U}{\xi' \eta'} = 0$ for any 
    $\xi,\xi' \in 2^n$ and $\eta,\eta' \in 2^m$ with $\xi \neq \xi'$.
    With this strengthening, it is easily seen that the 
    interpretation of {\rm\lstinline{if}} is retracted by its Hermitian 
    conjugate.
    Since the interpretation is a square matrix, it must be unitary.
    The other cases are straightforward.
  \end{proof}
\end{lemma}

Now, let us connect this two-layer probability semantics to the more 
established density matrix representation (for which see 
\cite{NielsenChuang2011QuantumComputationQuantum}).
A two-layer state is just an ensemble, so there is a unique density 
matrix reproducing its measurement statistics: 
$\sum_j p_j \sum_k\sum_{k'} q_{j,k} q_{j,k'} \ketbra{k}{k'}$.\footnote{The 
  $q_{j,k'}$ needs to be conjugated when generalizing to complex 
  amplitudes.
  The trick of swapping indices like $q_{j,k'} \mapsto q_{k',j}$ to express 
  conjugation common in superficially similar sums does not work, 
  since the $q$'s are not entries of a unitary matrix but a 
  coefficient indexed by indices with different ranges.}
The only difference is that the two-layer state remembers which 
ensemble it arose from, whereas the density matrix conflates 
observationally indistinguishable ensembles, similar to the 
relationship between $\beta\eta\delta$ equality and observational equivalence 
in PCF.

By \cref{thm:unitarity}, the two-layer probability semantics for 
computational statements are of the form 
$[(p_j, \sum_k q_{j,k}\ket{k})]_j \mapsto [(p_j, \sum_k 
q_{j,k}U\ket{k})]_j$ for a unitary $U$, which is simulated by the 
completely positive map $\rho \mapsto U \rho U^\dagger$ in the 
density matrix representation.
Allocation by \lstinline{new} can be understood along similar lines 
using the isometry $I_{2^n}^{2^{n+k}}$ in place of $U$.
One can check that the formula for measurement from 
\cref{sec:measurements} is simulated by projective measurements, 
which map $\rho \mapsto \sum_{r \in 2^n} P_r \rho P_r^\dagger$ where 
$P_r$ is a projector that projects to the subspace where the measured 
$n$ bits have value $r$.
Thus, we get:

\begin{theorem}
  Core QPPL semantics is implementable by unitaries, initialization, 
  and measurement.
  More precisely, the translation to density matrices is 
  computationally adequate (in the sense of reproducing measurement 
  statistics), and the translated semantics is implementable by those 
  operations.
\end{theorem}

Conversely, core QPPL can implement computationally universal sets of 
quantum gates.
An easy choice is Toffoli ($z \xorassign x \land y$) and Hadamard 
($\qrand{x}$) \cite{Aharonov2003SimpleProofToffoli}.
Therefore:

\begin{theorem}
  Core QPPL is computationally universal.
\end{theorem}

\begin{remark}
  There are several notions of universality of quantum gate sets.
  Strict universality means that any unitary matrix can be 
  approximated.
  Computational universality \cite{Aharonov2003SimpleProofToffoli} is 
  similar but allows quantum states to be encoded in a non-standard 
  way.
  Since we chose to present QPPL as a real-only formalism, we 
  established computational universality over the encoding of complex 
  amplitudes by real ones.
  If strict universality is desired, one can generalize to complex 
  amplitudes and add a variant of \lstinline{qnegate()} that 
  multiplies $i$ to the current world's amplitude, allowing to 
  implement the strictly universal gate set of Hadamard, CNOT, and 
  $S$.
\end{remark}

\section{Related Works}

Apart from Aaronson 
\cite{Aaronson2004QuantumMechanicsIsland,Aaronson2013QuantumComputingDemocritus} 
and Mu and Bird \cite{MuBird2001FunctionalQuantumProgramming} 
mentioned in the introduction, the view of quantum computation as 
non-deterministic transitions weighted by possibly-negative 
probabilities features prominently in the textbook proof of 
$\text{BQP} \subseteq \text{PSPACE}$.
The idea that negative probabilities is the essence of quantum 
computing goes all the way back to the foundation of the field: 
Feynman identifies them as the primary feature of quantum processes 
that cannot be efficiently simulated on classical computers 
\cite{Feynman1982SimulatingPhysicsComputers}.
The computational universality of real-amplitude gates 
\cite{Aharonov2003SimpleProofToffoli} can be seen as another 
vindication of the primacy of negative probabilities.
So the idea that quantum computing is characterized by negative 
probability is quite pervasive, but few have tried to distill a 
programming paradigm based on it.

The pseudocode conventions by Knill 
\cite{Knill1996ConventionsQuantumPseudocode} often gets credit as the 
first precursor to quantum programming languages.
Van Tonder \cite{vanTonder2003LambdaCalculusQuantum} was among the 
first to investigate functional quantum programming, a key feature of 
which was the use of linear types to rule out replication of 
variables, motivated by the no-cloning theorem.
It was Altenkirch and Grattage 
\cite{AltenkirchGrattage2005FunctionalQuantumProgramming} who pointed 
out that one could rather control discarding instead of copying, as 
the latter could be understood as sharing.
This idea was a part of the inspiration for QPP, which makes it 
obvious that data can be copied world-by-world.
A plethora of other quantum programming languages have been proposed 
\cite{SandersZuliani2000QuantumProgramming,%
  Omer2003StructuredQuantumProgramming,%
  JorrandLalire2005QuantumPhysicsProgramming,%
  LampisEtAl2008QuantumDataControl,%
  SelingerValiron2009QuantumLambdaCalculus,%
  YingEtAl2009AlgebraQuantumProcesses,%
  GreenEtAl2013QuipperScalableQuantum,%
  WeckerSvore2014LIQUiSoftwareDesign,%
  SvoreEtAl2018EnablingScalableQuantum,%
  PechouxEtAl2020QuantumProgrammingInductive,%
  BichselEtAl2020SilqHighlevelQuantum}, but all rest on the 
assumption that quantum programs should manipulate quantum data, 
whose finicky properties the programmer should know from studies 
elsewhere.

In classical computing, ``probabilistic programming'' has come to 
mean writing code that samples from infinite or even continuous 
distributions, possibly employing a post-selection construct that 
filters the samples for desired features 
\cite{vandeMeentEtAl2021IntroductionProbabilisticProgramming}.
The resulting artifact is often not readily executable with 
reasonable resources and may have subtle semantics due to 
singularities \cite{Jacobs2021ParadoxesProbabilisticProgramming}.
It is used more like a formal model of a stochastic process 
supporting statistical inference than an executable program.
QPP can perhaps be adapted for similar modeling and analysis of 
quantum processes, though whether that can be useful is unknown.


\section{Conclusion}
\label{sec:conclusion}

We proposed a new quantum programming paradigm, QPP.
It views quantum programming as probabilistic programming with 
negative probabilities, doing away with the concept of qubits and 
putting all the quantum weirdness into the quantum non-determinism 
effect.
This makes quantum programming more accessible by systematically 
providing computationally intuitive metaphors not involving advanced 
physics.
Throughout the paper, we talk only of computationally relevant 
concepts like the desire to control destructive interference for 
speed gains, instead of physical concepts like limitations on 
ripple-effects from measurements due to relativistic no-signaling 
principles, or intuition-free linear algebraic concepts like 
eigenvalues of Hermitian operators.

This paper only fleshed out the core ideas of QPP and its conceptual 
advantages over qubit-based programming.
Much work remains on growing this into a full-fledged, practical 
programming language, adding such things to QPPL as: more advanced 
data types like numbers and inductive data types, user-defined 
functions, iteration, and more sophisticated handling of allocation.
Higher-level intuitions that aid algorithm design are also wanted.

Another topic of future research is communication.
QPP views a single lump of qubits as one quantum non-deterministic 
computation.
Independent quantum computers, or mutually unentangled chunks of a 
single quantum computer's memory, can be naturally understood as 
multiple such computations.
Formalizing communication between such lumps as a merger of multiple 
non-deterministic computations would not only help to model 
communication protocols but also reconstruct the concept of qubits 
and quantum data, equipped with a perspective for better 
understanding and manipulating what's inside.

\begin{credits}
  \subsubsection{\ackname} \ifanonymized (anonymized for review)
  \else
    We thank Hideaki Nishihara and Akira Mori 
    for insightful comments on early drafts of this paper.
  \fi

\subsubsection{\discintname}
The author has no competing interests to declare that are relevant to 
the content of this article.
\end{credits}
%
%
%
\bibliographystyle{splncs04}
\bibliography{local.bib}
\end{document}